\documentclass[preprint,12pt]{elsarticle}

\usepackage{amssymb}
\usepackage{longtable,tabu}
\usepackage{subfigure}
\usepackage{url}
\usepackage{multirow}

\usepackage{hhline} 
\usepackage{supertabular}

\newcommand\arraybslash{\let\\\@arraycr}

\journal{XXXX}

\begin{document}

\begin{frontmatter}

\title{Slime mould: the fundamental mechanisms of biological cognition}

\author[1]{Jordi Vallverd\'{u}}
\author[1]{Oscar Castro}
\author[3]{Richard Mayne}
\author[2]{Max Talanov}
\author[6]{Michael Levin}
\author[5]{Frantisek Balu\v{s}ka}
\author[4]{Yukio Gunji}
\author[8]{Audrey Dussutour} 
\author[9]{Hector Zenil}
\author[3]{Andrew Adamatzky}

\address[1]{Department of Philosophy, Universitat Aut\`{o}noma de Barcelona, Catalonia}
\address[2]{Kazan Federal University, Kazan, Russia}
\address[3]{Unconventional Computing Centre, University of the West of England, Bristol, UK}
\address[4]{Waseda University, Tokyo, Japan}
\address[5]{Institute of Cellular and Molecular Botany, University of Bonn, Germany}
\address[6]{Allen Discovery Center, Tufts University, Medford, MA,USA}
\address[8]{Universite Paul Sabatier, Toulouse, France}
\address[9]{Algorithmic Dynamics Lab, SciLifeLab, Karolinska Institute, Stockholm, Sweden}

\begin{abstract}
The slime mould {\it Physarum polycephalum} has been used in developing unconventional computing devices for in which the slime mould played a role of a sensing, actuating, and computing device. These devices treated the slime mould rather as an active  living substrate yet the slime mould is a self-consistent living creature which evolved for millions of years and occupied most part of the world, but in any case, that living entity did not own true cognition, just automated biochemical mechanisms.  To ``rehabilitate'' the slime mould from the rank of a purely living electronics element to a ``creature of thoughts'' we are analyzing the cognitive potential of {\it P. polycephalum}. We base our theory of minimal cognition of the slime mould on a bottom-up approach, from the biological and biophysical nature of the slime mould and its regulatory systems using frameworks suh as Lyon’s biogenic cognition, Muller, di Primio-Lengeler\'{s} modifiable pathways, Bateson's  ``patterns that connect'' framework, Maturana’s autopoetic network, or proto-consciousness and Morgan’s Canon.
\end{abstract}


\end{frontmatter}

\newpage 

\tableofcontents

%\newpage

%%
%% Start line numbering here if you want
%%
% \linenumbers

%% main text

\graphicspath{{figs/}}

%\section{Slime Mould as a Fundamental Minimal Cognitive Entity}

\section{Introduction}

Classic approaches to cognition were based on human beings, or sometimes, included some great apes or close mammals, leaving unattended a long list of living entities~\cite{waal}. This biased anthropomorphic perspective had a second problem: it blocked pathways to an evolutionary and naturalistic study of cognition~\cite{gods}. It is easy to find several studies on minimal examples of cognition  among prokaryotes \cite{muller2001contributions}, being protists a perfect example of it. Of them,  ameboides (Myxomycetes) are the best example of the power of such cognitive studies studies, because they show us the connection bridge between  unicellular ({\it Physarum}) to pluricellular ({\it Dictyostelium}) living systems. Thus, {\it Physarum polycephalum} shows to be an extremly intelligent system which offers unique ways for the understading of the emergence of complex behaviours and cognitive strategies \cite{dussutour2010amoeboid,shirakawa2011associative,mori2013cognition}. 
The mapping of such minimal cognitive elements in Slime mould makes also possible the analysis of two different but fundamental ideas: first, to identify the computational nature of basic cognitive processes, which in our case offer a good example of a reliable Kolmogorov-Uspensky biomachine; this biocomputational approach offers a naturalistic way to explain the intricacies of informational processing from scratch to the supervenience of proto-consciousness. This process can be explained without relying on any anthropomorhic bias thanks to implementation of the Morgan's Canon.As a consequence, slime mould offers a unique and special biological framework for the identification of the basic biocomputations that make possible cognition and pave the way for the emergence of consciousness (understood as meta-level of informational processing). In the next section we describe in more detail the value of such selection for the study of cognition.
% * <adamatzky@gmail.com> 2017-11-25T09:27:06.819Z:
%
% > it blocked or ra counter to a
%
% ^.

\section{Minimal Cognition: the Bottom-up Approach to Cognition}

After the long and intense analysis of human, primate and mammals cognition, the necessity of explaining their evolutionary and basic functional properties led to the interest for minimal forms of cognition. This minimal cognition interest was then focused into plants
\cite{minimal}  \cite{calvo} and the fundamental chemical mechanisms that rule their behaviour \cite{chemical}. Fundamental  aspects of biological systems were detected: they process, store and process information, a kind of computation \cite{free} \cite{koseska}. Other studies have also analyzed the cognitive skills of bacteria \cite{jacob}, \cite{shapiro}, or \cite{jacob2}. Despite some attempts of defining a minimal structure for the existence of cognition \cite{barandiaran}, we strongly disagree with the idea that emotional and conscious spheres are necessarily and uniquely dependent to neurodynamic processes \cite{moreno}. The principles that allow, justify and sustain consciousness must rely at a lower level, and other mathematical approaches have provided simple informational mechanisms able to be performed by slime moulds as the basis of data integration and conscious experience. Despite some very interesting attempts at looking for the phylogenetic emergence of consciousness \cite{cabanac}, most of the studies elude the analysis of fundamental aspects which could explain a naturalistic and scalable explanation of consciousness increase in complexity. Some general analysis about this new approach has been done by \cite{vallverdu2017} and will be explained with more detail into next sections. In no way are we embracing panpsychist or panexperientalist approaches, as could be misleadingly inferred. At the same time we wish not to dilute our research into unending philosophical debates. Slime moulds know and gather knowledge via their specific proto-consciousness.

Taking into account these elements we wish to descend even more into the tree of life and select slime mould as a very minimal cognitive system \cite{slime1}. Despite of the very simple cognitive architecture, slime mould are able to perform an incredible list of cognitive tasks as well as to be creative \cite{creativity}, considering this skill as an emergent property of their genomic structure, as it works for other living systems \cite{bacterial}.
As noted by \cite{slimebook}, the slime mould shows outstanding abilities to adapt its protoplasmic network to varying environmental conditions in which it can solve tasks of computational geometry, image processing, logic and arithmetic if and only if data are represented by binary configurations of attractants and repellents.  We speculate that in slime mould cognition is embedded with computation in morphological patterns of protoplasmic networks updated locally similarly to a computational process travelling in disordered dynamically changing graphs, or storage structures, of Kolmogorov-Uspensky machines~\cite{kolmogorov1953concept,uspensky1992kolmogorov,gurevich_1988,blass2003abstract}.

\section{Defining the Nature of Slime Mould} 

%Rich: Clarified paragraph to include all 3 slime mould branches and put in proper nomenclature
The slime moulds are a polyphyletic group of protistic organisms existing in three major categories: `true', acellular or plasmodial slime moulds of class {\it Myxogastria}, cellular or pseudoplasmodial slime moulds of class {\it Dictyosteliida} and the unicellular microscopic slime moulds of class {\it Protosteliales} \cite{Stephenson1994}. Of the former two varieties, the first is a multinucleate single cell and the second is a multicellular complex. Therefore, all slime moulds could be tested to minimal cognition for proto-consciousness.

The basic regulatory functions (as metabolic ones) can explain the emergence of cognitive mechanisms \cite{bich}, and therefore both homeostatic and homeodynamic perspectives can allow the existence of a minimal cognition. But in no case proto-consciousness can only be considered as the summatory of different minimal cognitions, instead of the result of the emergency of entrangled multi-taxis or multy-tropisms functioning in coherence or decoherence playing.

%"minimal cognition" for categorical view of processes and "proto-cognition" in molecular level, as molecular memory, switch behaviour, and so. Is possible to say "micro-cognition" in bacterial colonies adaptive behaviour?. But in slime molds is necessary rethink it.

Acellular slime mould \emph{P. polycephalum} has a sophisticated life cycle~\cite{stephenson1994myxomycetes}, which includes fruit bodies, spores, single-cell myxamoebae and the plasmodium, a multinucleated syncytium \cite{Mayne2016biology}. The plasmodium is a coenocyte: nuclear divisions occur without cytokinesis.  The plasmodium is a large cell. It grows up to tens of centimetres when conditions are good.
The plasmodium consumes microscopic particles, bacteria and flakes oats. During its  foraging behaviour the plasmodium spans scattered sources of nutrients with a network of  protoplasmic tubes. The plasmodium optimises it protoplasmic network to cover all sources of nutrients, stay away from repellents and minimise transportation of metabolites inside its body. The plasmodium's ability to optimise its shape~\cite{nakagaki2001path} attracted attention of biologists, then computer scientists~\cite{adamatzky2010physarum}  and engineers. Thus the field of slime mould computing was born. 

So far, the plasmodium is the only  stage of \emph{P. polycephalum}'s life cycle useful for computation. Therefore further we will use word `Physarum' when referring to  the plasmodium. Most computing and sensing devices made of the Physarum explore one or more key features of the Physarum's physiology and behaviour: 
\begin{itemize}
\item the slime mould senses gradients of chemo attractants and repellents~\cite{durham1976control, ueda1976chemotaxis, rakoczy2015application}; it responds to chemical or physical stimulation by changing patterns of  electrical potential oscillations~\cite{ridgway1976oscillations, kishimoto1958rhythmicity} and protoplasmic tubes contractions~\cite{wohlfarth1979oscillatory, teplov1991continuum};  
\item it optimises its body to maximise its protoplasm streaming~\cite{dietrich2015explaining}; and, 
\item it is made of hundreds, if not thousands, of biochemical oscillators~\cite{kauffman1975mitotic,Mayne2016oscillators} with varied modes of coupling~\cite{grebecki1978plasmodium}.  
\end{itemize}

Slime mould can be seen, if we think at a very simplified level, as a reaction-diffusion excitable system encapsulated in an elastic growing membrane \cite{adamatzky2007physarum}. The behaviour of the Physarum is governed by an ensemble of thousands of biochemical oscillator, who set local clocks and control peristaltic activity and growth via propagation of actin polymerisation waves and calcium waves. A coupling between distant parts of Physarum via waves of electrical potential is manifested in oscillations of the electrical potential.

\section{What Does a Slime Mould Know?} 
%In this section we will explain the set of cognitive informational actions performed by slime mould during their acquisition steps.

%\begin{figure}[ht]
%    \centering
%    \includegraphics[width=0.6\textwidth]{gradients}
%    \caption{The world as seen by slime mould. Sites with highest concentration of attractants are blue, repellents are red}
%    \label{fig:gradients}
%\end{figure}

%\begin{figure}[ht]
%    \centering
%    \includegraphics[width=0.6\textwidth]{PhysarumSalt}
%    \caption{Physarum propagating on a nutrient agar gels avoids domains with high concentration of repellent. In this particular experiment we chosen sodium chloride as repellent. Locations of the sodium chloride are shown by stars. See details in \cite{adamatzky2010routing}.}
%    \label{fig:salt}
%\end{figure}

Living systems process information in order to react  to the environment and to be able to survive or to transmit their own informational structure \cite{terzis2011information}. We can define this process as ``knowledge'', because the living system does not produce automated responses,like a thermostat can do, but it imply certain  evaluation.

Physarum lives in the world of gradients, 
%(Fig.~\ref{fig:gradients}, 
concentrations of attractants and repellents, and consequently its behavioural responses are a direct consequence of their interaction with these gradients. For example: while propagating on a substrate the slime mould avoids domains with concentrations of repellents exceed a threshold, 
%(Fig.~\ref{fig:salt}), 
see e.g. \cite{adamatzky2010routing}.

\section{Modifiable stimulus-response pathways from an autopoietic perspective [base of all behaviours]}

We must admit that one of the key roles for the development of a living system is its operational autonomy according to the organism's own internal program \cite{rosen1991life,rosen2013optimality}. 

This determines the basic actions with which the system reacts to stimuli from the 
environment--stimulus acting on the system through certain internal environment structurally coupled receptors. In fact, Maturana and Varela \cite{maturela} determined their autopoietic theory considering that the problem of autonomy had to be reduced to its minimal form, in the characterization of the basic living unit.

An autonomous system that is self-maintained (as it is autopoietic) generates both development paths and coherent organization, allowing that we describe ii as an agent. But the relationship between the agency and cognition, within the framework of autonomy, is still precarious. With an autonomous system is self-maintained, offering a hemodynamic balance is not sufficient reason for the emergence of a discriminatory semiosis and therefore a minimum cognitive principle.

In short, the basic self-maintenance networks of metabolism, capable of certain forms of adaptive responses do not show those skills necessary for the emergence of minimal cognition \cite{castrofilosofia}. A more complex organization, which gathers mechanisms that go beyond this basic metabolism, is necessary. That is, that the mechanisms of internal compensation of interactions with the immediate environment are responsible for the emergence of a biosemiotics discrimination of second level --- macromolecular --- allowing both the reception distinguished between what is noise which is significant for the system agential autonomous.

We also take into account the generation of autopoietic subsystems when to explain the minimum domains of perceptual activity in a structurally stable autopoietic organization. These establish a new criterion of interdependent autonomy within a hierarchy in the organization of living things. Own cellular components are established as autopoietic interdependent subsystems in a hierarchical level lower than the cell. Subsystems comprising a minimum subvenient enactivity both morphologically and physiologically, exerting a specific regulatory function. We can say, in the words of Leonardo Bich and Alvaro Moreno \cite{bich, moreno}  that are specific regulatory subsystems.

It is understood as regulation as the capacity of living beings of internally compensates disturbances \cite{moreno}. Exogenous environmental phenomena to organism or disturbances are managed by the agential systems that confer significant and therefore an intrinsic biosemantics to dynamic patterns of transduced self-regulation. In the words of Piaget, it is assimilation and an accommodation \cite{piaget} to the operational closure of the agent to overcome the binomial ``stimulus -- response''. That is, the interaction between an organism and the environment through the adaptation ontogenetic considered an assimilation of external influences; it involves an internal self-regulation, or accommodation, and thus a restructuring of the system. This provides a dynamic and cyclical view of self-regulation: endogenous action $\rightarrow$ interaction with Umwelt $\rightarrow$ self-regulatory disturbance compensation $\rightarrow$ internal re-balances, etc \cite{castro2009jakob}.

To achieve effective regulatory an interdependent control autopoietic subsystem is necessary. But that is sufficiently independent of the dynamics of the processes controlled, and can be changed without interrupting these processes; but also it is able to be linked to parts of the control mechanism of the system --- a regulated subsystem --- in order to be able to modulate its operations \cite{bich}. More specifically, the appearance of a control subsystem implies that the organism itself generates a set of decoupled processes from the dynamics of their constitutive system. The dynamic decoupling between the regulator and the regulated constituent subsystems means that both --- although they are correlated through the system --- are integrated, working at different intrinsic rhythms. That is, the regulation components are produced and maintained by the activity of the constituent subsystems regulated, and said activity, in turn, is modulated by the regulator.

As occur in the TCST system of {\it Escherichia coli} bacteria \cite{stock2000two} this happens when the function of the regulation subsystem is neither specified nor directly determined by the metabolic activity of constituent subsystems regulated: that is, is ``stoichiometrically free'' latter \cite{griesemer}, so no biochemical balance between them. More specifically, the fundamental idea of this form of decoupling is that activation and operation of the subsystem control is not directly dependent on its concentration (or concentration variation), ie of its production by the constituent subsystems regulated, even although these subsystems ensure their presence in the system. In contrast, activation control subsystem is caused by environmental disturbances, and operations depend on their internal organization and structure of its functional components. So the regulator subsystem can work operationally differently from the constituent subsystems regulated, and can, in principle, act as a regulator controller of regulated constituent subsystems \cite{bich}.

\subsection{Recognition in regulators subsystems}

In this organizational architecture, the functional role of a regulator subsystem is modular basic constituent network, switching between different metabolic systems available in the system in relation to changes in environmental conditions. It does so in such a way that new metabolic/constituent regimes engendered by regulatory switches must be able to cope with the new environmental conditions, expanding the range of disturbances or stimuli to which the system can respond quickly and efficiently, and enrich the field of dynamic functional behaviours available.

The crucial thing is that the organism endowed with regulatory systems reacts in a new way, do things according to what it recognizes and distinguishes in its interactions with ``Umwelt'' of the autopoietic regulator subsystem. In fact, the ``recognition'' in fact, is a complex process, since the specific characteristics of the interactions with the Umwelt are responsible for triggering the subsystem regulations cannot directly drive the system response, as in the case of responses basic network. In fact, a disruption active certain regulation subsystem, which in turn modulates the basic constituent network so as to meet the specific environmental characteristic that triggered the regulatory response.

A couple of examples applicable to myxomycete {\it P. polycephalum} would be: 
\begin{itemize}
\item that the organism is powered by a new source of food richer in nutrients that can win more energy, even if you decide to choose the path worst environmental condition which should go with a poor nutrient food source \cite{takagi}. 
\item that transforms their morphology --- plasmodium to sclerotium --- to preserve them in a hostile environment lacking food, minimizing your breathing \cite{seifriz} maximizes its longevity \cite{seifriz2, gehenio, jump}. 
\end{itemize}

In other words, the fact of recognition is a consequence of the specific nature of the disturbance and internal regulatory organization system.

\subsection{Recognition and equivalence classes}

In such scenario an environmental disturbance becomes a specific and recognizable interaction due to the nature of the relationship with the subsystem controller. Regulator subsystem is responsive to the disturbance, in the sense that endogenously establishes the equivalence classes in their environment with respect to these specific changes, according to how variations activate the controller subsystem to trigger regulatory action. These equivalence classes not consist externally associations between the disturbances and the results of the changes brought about in the system as a whole, as in the previous case. Rather, they are the results of the evaluation (activation control subsystem further regulatory action on the constituent subsystems regulated) operated by their regulation subsystem that achieve endogenous and functional significance for the system.

Therefore, beyond the debate about whether cognition is coextensive with life, it is clear that regulation becomes a necessary condition for the appearance of cognition, considering as an essential aspect of cognition that the knower should be capable of transforming an external influence on an adaptive integration or, in cognitive terms, in a meaningful interpretation~\cite{moreno}. And this requirement is essential for cognition is carried out only by regulation rather than by self-production basic biological and self-maintenance (i.e. minimal autopoietic systems), as some of the defenders of the tautology of life and cognition have argued. 

\section{Significant Regulation}

A system with normativity capacity, in fact, is able to do things according to what distinguishes (what the regulation is sensible subsystem). Consequently, interactions with the environment become more than a source of confusing noise, but become a world generated endogenously (naturalized) of meanings --- An Umwelt~\cite{von, castro2011principles}: interactions functionally become ``significant'' for the system itself, without being to an outside observer.

Here the term ``meaning'' as a synonym for``functional for the system'' (see \cite{barandiaran}) is used. More specifically, a source of disturbance makes sense when you can distinguish by the regulatory subsystem and this distinction has an operational effect on the system. Through the action of the regulatory system subsystem modulates its own activity (in the constituent subsystems regulated) on the basis of this distinction is as contribute to their own self-maintenance.

\subsection{Electrical activity}

Another example  of normative regulatory system an be set in the regulation of 
Ca$^{2+}$ for activation or inhibition of myosin allowing oscillations of current flow for displacement plasmodium. Here Ca$^{2+}$ acts as an inhibitor of the oscillations of the actin-myosin, but in the opposite direction of the existing inhibitory activity in the process of contraction of animal muscle cells. The absence of Ca$^{2+}$ in neuromuscular processes inhibits contracture activity while in plasmodia; inhibition increased by Ca$^{2+}$ \cite{nakamura} arises.
Ca$^{2+}$ but is a powerful oscillator in the presence of cAMP in the environment of plasmodium \cite{smith}.

More broadly, the extensive utilization of bioelectric phenomena by Phy\-sarum \cite{adamatzky2011electrical,halvorsrud1995patterns,achenbach1981ionic,kishimoto1958rhythmicity}  dovetails with the broader use of ionic signaling in morphological computation throughout metazoan. Indeed, ion channel and neurotransmitter hardware, which underlie the software of cognition in animal brains, are evolutionarily ancient and pre-date multicellularity \cite{levin2006minds,keijzer2013nervous,moran2015evolution,liebeskind2011evolution}.  For a review of cognitive approaches to the decision-making and memory in non-neural metazoan cells and tissues in the context of embryogenesis, regeneration, and cancer suppression see \cite{baluvska2016having,pezzulo2015re,mathews2017gap}. The reading of bioelectric states of Physarum during learning and problem-solving, to achieve a kind of ``neural'' decoding, is an on-going effort at the frontier this field.

\subsection{Regulatory subsystems independent of metabolic processes}

In the same way that the phenomenon of chemotaxis for the bacteria {\it E. coli} seen, also myxomycetes offer specific chemotaxis, both multicellular pseudoplasmodium {\it Dictyostelium discoideum} \cite{bonner, mari, song} and the acellular plasmodium {\it P. polycephalum} \cite{keller2, knowles, ueda1976chemotaxis}

But in the case of taking the phenomenon of chemotaxis as a condition for cognitive process requires as \cite{moreno} explains that part of the regulatory subsystems are independent of metabolic processes. This case is a complex form of driven stability taxis, achieved through the coupling of two subsystems and indirect feedback through the environment. The movement here is autonomous and depends on the internal differentiation of the organization and activity of inherent self-maintenance of the cell. Additionally, this behaviour is functional in the sense that contributes to the maintenance of conditions of existence of metabolism and consequently of the entire system. However, in this example, although the system is capable of complex viable behaviours, the system responds as a network together without distinguishing the specific of their interaction with the environment, evaluating them and modulating action accordingly. In this case, the environment is only a noise source that disturbs the metabolism, and the behavioural response is filtered through the latter.

This chemotactic mechanism involves at least three molecular complexes:
\begin{itemize}
\item a receiver subsystem, which is activated by environmental effectors (chemoattractants and chemorepellents); 
\item  a conductive tip in the case of pseudopod plasmodia and pseudoplasmodia, and 
\item  a group of molecules, and macromolecular filaments which act as a bridge between the two. Usually, the latter are activated molecules through non-covalent post-transducer modifications, such as phosphorylations cascades.
\end{itemize}

The distinguishing feature of this complex case of chemotactic behaviour is that the system, thanks to the action of a subsystem control, is able to modulate their behaviour based on specific interactions with the environment, which acquires a specific operational meaning for the system, so that the latter is able to change their behaviour accordingly. This is made possible by the fact that the regulation subsystem is decoupled from the other. This means that the regulatory subsystem --- in the case of plasmodia {\it P. polycephalum} would be reaction-diffusion pattern (\cite{keller} through the filaments and myosin microtubules. It is the stimulus that comes from attractants or repellants (external) requiring adapts the material to their spatial behaviour. The specific mechanism used is the diffusion of attractants or repellants in the environment. The presence of these stimuli in the periphery of the material provides the impetus for morphological adaptation \cite{mayne}.

This decoupling actually introduces a new degree of freedom in the system, one or more new variables in the controller subsystem which does not depend directly on the constitutive regime and therefore may be sensitive to something different than the internal state of the system: in this case a characteristic of the environment. This characteristic takes an important role for the entire system, and subsystem of decoupling achieves a functional role when the effect of its regulatory action occurs, caused by the disturbance and contributes to system maintenance. From this relatively elementary chemotactic mechanism, then, a variety of more complex tactical behaviours can be implemented \cite{bich, moreno}. What is common to them is that cannot be treated in terms of input-output, as if the behaviour of the system was expelled by the disturbance. On the contrary, they can be characterized in terms of endogenous generation of adaptive response by focusing on the internal organisation of the system and, above all, regulatory subsystems.

So we conclude chemotaxis of myxomycetes --- such as bacterial --- represent a clear comparative example of how certain fundamental capabilities for the origin of cognition may arise in the minimum living systems, and how this is possible only through the action of regulatory mechanisms. Only in the presence of regulation that specific disturbances acquire a meaning for the system. This is a biosemiotic foundation of the basics of cognitive minimum principles, emerging regulatory system factors acquire those minimum principles offered in cognitive biology \cite{maturana}\cite{bateson}\cite{maturela}\cite{muller}\cite{lyon}\cite{lyon2}.

\section{Minimal Cognitive Principles in Myxomycetes. Cognitive biology as framework}

From the Darwinian theory of species each organism is evolving from a common root. This is very important as the foundation of a common root of the minimal cognitive principles that underpin the higher emergent developments. For this reason, three additional elements are required in cognitive biology:
\begin{itemize}
\item the study of cognition in a species of organism is useful, through contrast and comparison, for the study of the cognitive capacities of another species \cite{spetch2006comparative}; 
\item to proceed with more complex cognitive systems, it is useful to start from the simplest organisms \cite{baluvska2009deep}, and 
\item For a better understanding of the nature of cognition, the greater the number and variety of species studied in this sense \cite{lyon2015cognitive}.
\end{itemize}
While cognitive science strives to explain human thought and the conscious mind, the work of cognitive biology focuses on the most fundamental process of cognition of any organism \cite{goodwin1978cognitive}.

Cognitive biology is, therefore, a new approach to biological paradigms about cognition that need a coherent significantly contextualisation, both to experimental data and to a semantic development of biology.

Ladislav Kovac of the Department of Biochemistry and Genetics of the Faculty of Natural Sciences of the Comenius University in Bratislava states in \cite{kovavc2000fundamental} that:

\begin{quote}
``Cognitive biology is more a reinterpretation of existing data than a research program that offers new experimental approaches to old problems.''
\end{quote}

and adds:

\begin{quote}
``Cognitive biology aims at a synthesis of data from various scientific disciplines within a single framework of conceiving life as the epistemic unfolding of the universe (the epistemic principle). According to evolutionary epistemology, it considers biological evolution as a progressive process of accumulation of knowledge. Knowledge manifests itself in the constructions of organisms, and the structural complexity of the constructions that carry the incorporated knowledge corresponds to its epistemological complexity. In contrast to evolutionary epistemology, cognitive biology is based on the assumption that the molecular level is fundamental to cognition and adheres to a principle of minimum complexity, which states that the most efficient way to study any trait of life is by studying it at the simplest level in which it occurs.''
\end{quote}

For this reason the interest in a new reformulation of the concept of cognitive process and intelligence requires a deeper understanding of the behaviour of living systems, where cognition does not depend on the existence of a nervous system that channels it, but on functional circuits which allow the minimal perception of the surrounding environment and its biosemiotic processing in vivo \cite{kull2009theses}.

\section{Minimal Cognition Principles Illustrated by Tables of Conceptual Relationships}

\subsection{Pamela Lyon's Table}

Pamela Christine Lyon of the University of Adelaide in Australia argues that cognition is a natural biological phenomenon of maximum access, like other natural biological phenomena: by studying simple model systems (such as bacteria, myxomycete) can be understand the bases and then expand to more complex examples (from bees and ants to hominids). Lyon develops a biocognition (as says van Dujin \cite{van2012biocognitive}) from a biogenic framework, whose approach tries to anchor the concept of cognition itself in biology. In this approach, cognition is primarily a form of biological adaptation that confers certain selective advantages specific to organisms by allowing them to cope with environmental complexity.

Pamela Lyon \cite{lyon} synthesizes in ten biogenic principles the development that Kovac \cite{kovavc2000fundamental} made based on the principles of cognitive biology. This synthesis illustrates the depth of the bases with which they have been established in the school of cognitive biology of Adelaide and which are listed as follows:
\begin{enumerate}
\item Continuity: Complex cognitive abilities have evolved from the simplest forms of cognition. There is a continuous line of significant descent. (Do not rule out the appearance of new capabilities with greater complexity).
\item Control: Cognition directly or indirectly modulates the physical-chemical processes that constitute an organism.
\item Interaction: Cognition facilitates the establishment of reciprocal relations of causality with the characteristics of the environment, giving rise to exchanges of matter and energy that are essential for the continuous persistence of the organism, well-being or reproduction.
\item Normativity: Cognition refers to the (more or less) continuous assessment of the needs system relative to prevailing circumstances, the potential for interaction, and whether the current interaction is working or not.
\item Memory: Cognition requires the ability to retain information for a period of time greater than zero.
\item Selectivity: Because an organism is able to interact profitably with some, but not all, properties of the environment, cognition involves the differentiation of some states from the affairs of other states of affairs.
\item Valencia: In relation to the needs of the body and/or experience different environmental properties will be invested with different degrees of strength or importance, both positive and negative.
\item Anticipation: Cognition is intrinsically future oriented (what happens next?) And therefore predictive.
\item Reduction of randomness: Cognition is an important mechanism that biological systems reduce and modulate the influence of random perturbations on their functioning and, therefore, are resistant to disturbances.
\item Interdependence: The biochemical pathways that innervate cognition are intimately linked to those of other biological functions, making delimitation difficult and largely a function of explanatory objectives. (The map is not the territory).
\end{enumerate}

To find the minimum processes of cognition in cognitive biology we must take into account the production and processing of communication signals for the behaviour of the group and types of interconnection between individual single units (cells).
\newpage

\begin{flushleft}
\tablefirsthead{}
\tabletail{%
\hline
\multicolumn{3}{|r|}{\small\sl continued on next page}\\
\hline}
\tablelasttail{}
\tablehead{%
\hline
\multicolumn{3}{|r|}{\small\sl continued from previous page}\\
\hline
}
\tablelasttail{}
% supertabular
\begin{supertabular}{m{2.1754599in}|m{1.9740598in}m{1.8163599in}|}
\hline
\multicolumn{3}{|m{6.12336in}|}{\centering {\textbf{Minimum Cognitive Principles}}}\\\hline
\multicolumn{1}{|m{2.1754599in}|}{{\textbf{Pamela Lyon (2006)}}} &
\multicolumn{2}{m{3.8691597in}|}{{\textbf{{\it P. polycephalum} Behaviour}}}
\\\hline

\multicolumn{1}{|m{2.1754599in}|}{{\multirow{3}{*}{Continuity}}} &
\multicolumn{2}{m{3.8691597in}|}{{Tubulin isotypes relationship between {\it P. polycephalum} and different mammals, including humans \cite{clayton1980comparison}}}
\\\hhline{~--} 
\multicolumn{1}{|m{2.1754599in}|}{} & 
\multicolumn{2}{m{3.8691597in}|}{{Intracellular communication: reaction-diffusion model \cite{yamada2007dispersion}}}
\\\hhline{~--}
\multicolumn{1}{|m{2.1754599in}|}{} &
\multicolumn{2}{m{3.8691597in}|}{{Decision making and ``scatter'' or
state of indecision between different attractors for foraging \cite{takagi2007indecisive}}}
\\\hline

\multicolumn{1}{|m{2.1754599in}|}{{\multirow{6}{*}{Control}}} &
\multicolumn{2}{m{3.8691597in}|}{{Unit microtubules for migrating towards the conductive cores plasmodium area \cite{ueda2000microtubule}}}
\\\hhline{~--}
\multicolumn{1}{|m{2.1754599in}|}{}  &
\multicolumn{2}{m{3.8691597in}|}{{Decision making and ``scattor'' or
state of indecision between different attractors for foraging \cite{takagi2007indecisive}}}
\\\hhline{~--}
\multicolumn{1}{|m{2.1754599in}|}{} &
\multicolumn{2}{m{3.8691597in}|}{{Anticipation and memory periodic pulses \cite{saigusa2008amoebae}}}
\\\hhline{~--}
\multicolumn{1}{|m{2.1754599in}|}{}  &
\multicolumn{2}{m{3.8691597in}|}{{Distribution of nutrients \cite{dussutour2010amoeboid}}}
\\\hhline{~--}
\multicolumn{1}{|m{2.1754599in}|}{}  &
\multicolumn{2}{m{3.8691597in}|}{{Using his trail for spatial memory \cite{reid2012slime}}}
\\\hhline{~--}
\multicolumn{1}{|m{2.1754599in}|}{}  &
\multicolumn{2}{m{3.8691597in}|}{{A stimulus triggers the release of a signaling molecule, generating a feedback loop \cite{alim2013random}.}}
\\\hline

\multicolumn{1}{|m{2.1754599in}|}{{\multirow{5}{*}{Interaction}}} &
\multicolumn{2}{m{3.8691597in}|}{{Chemotaxis \cite{durham1976control}\cite{keller1970initiation}\cite{keller1971model}}}
\\\hhline{~--}
\multicolumn{1}{|m{2.1754599in}|}{} &
\multicolumn{2}{m{3.8691597in}|}{{Microtubule associated with chemotaxis \cite{ueda1994microtubules}\cite{ueda2005intelligent}}}
\\\hhline{~--}
\multicolumn{1}{|m{2.1754599in}|}{} &
\multicolumn{2}{m{3.8691597in}|}{{Irrationality making food choices \cite{latty2011irrational}}}

\\\hhline{~--}
\multicolumn{1}{|m{2.1754599in}|}{} &
\multicolumn{2}{m{3.8691597in}|}{{To leave chemicals in the environment, which are attractive to congeners to find food. Communication-based to cooperation and sociality \cite{vogel2015phenotypic}.}}
\\\hhline{~--}
\multicolumn{1}{|m{2.1754599in}|}{} &
\multicolumn{2}{m{3.8691597in}|}{{Fusion microplasmodia to form macroplasmodia in a percolation transition \cite{fessel2012physarum}.}}

\\\hline

\multicolumn{1}{|m{2.1754599in}|}{{Normativity}} &
\multicolumn{2}{m{3.8691597in}|}{{Rules for biologically inspired design adaptive networks
\cite{tero2010rules}}}
\\\hhline{~--}
\multicolumn{1}{|m{2.1754599in}|}{} &
\multicolumn{2}{m{3.8691597in}|}{{Comparative evaluation standards for decision making food
\cite{latty2011irrational}}}
\\\hhline{~--}
\multicolumn{1}{|m{2.1754599in}|}{} &
\multicolumn{2}{m{3.8691597in}|}{{Precise speed decision making \cite{latty2011structure}}}
\\\hline

\multicolumn{1}{|m{2.1754599in}|}{{\multirow{2}{*}{Memory}}} &
\multicolumn{2}{m{3.8691597in}|}{{Anticipation and memory periodic pulses \cite{saigusa2008amoebae}}}
\\\hhline{~--}
\multicolumn{1}{|m{2.1754599in}|}{} &
\multicolumn{2}{m{3.8691597in}|}{{Using his trail for spatial memory \cite{reid2012slime}}}
\\\hline

\multicolumn{1}{|l|}{{Selectivity}} &
\multicolumn{2}{m{3.8691597in}|}{{Decision making and ``scatter'' or
state of indecision between different attractors for foraging \cite{takagi2007indecisive}}}
\\\hline

\multicolumn{1}{|l|}{{\multirow{2}{*}{Valencia}}} &
\multicolumn{2}{m{3.8691597in}|}{{Unit microtubules for migrating towards the conductive cores
plasmodium area \cite{ueda2000microtubule}}}
\\\hhline{~--}
\multicolumn{1}{|l|}{} &
\multicolumn{2}{m{3.8691597in}|}{{Relationship of flow of electric current and polarization
associated with filaments and myosin microtubules, in turn coordinate current flow (streaming) and plasmodium oscillations for ``mobiligence'' \cite{mayne2015role}}}
\\\hline

\multicolumn{1}{|l|}{{Anticipation}} &
\multicolumn{2}{m{3.8691597in}|}{{Anticipation and memory periodic pulses \cite{saigusa2008amoebae}}}
\\\hline

\multicolumn{1}{|l|}{{\multirow{5}{*}{Reduction randomness}}} &
\multicolumn{2}{m{3.8691597in}|}{{Rules for biologically inspired design adaptive networks \cite{tero2010rules}}}
\\\hhline{~--}
\multicolumn{1}{|l|}{} &
\multicolumn{2}{m{3.8691597in}|}{{Reaction-diffusion model for streaming oscillation patterns \cite{yamada2007dispersion}}}\\\hhline{~--}
\multicolumn{1}{|l|}{} &
\multicolumn{2}{m{3.8691597in}|}{{Composition of a set of networks basic topologies \cite{nakagaki2004obtaining}}}\\\hhline{~--}
\multicolumn{1}{|l|}{} &
\multicolumn{2}{m{3.8691597in}|}{{Using his trail for spatial memory \cite{reid2012slime}}}
\\\hhline{~--}
\multicolumn{1}{|l|}{} &
\multicolumn{2}{m{3.8691597in}|}{{Detection stimulus intensity differences through the
Weber-Fechner Law \cite{mori2013cognition}}}
\\\hline

\multicolumn{1}{|m{2.1754599in}|}{{\multirow{2}{*}{Interdependence}}} &
\multicolumn{2}{m{3.8691597in}|}{{Plasmodium to sclerotia \cite{seifriz1939materialistic}, \cite{gehenio}, \cite{jump1954studies}}}
\\\hhline{~--}
\multicolumn{1}{|l|}{}  &
\multicolumn{2}{m{3.8691597in}|}{{Relationship of flow of electric current and polarization associated with filaments and myosin microtubules, in turn coordinate current flow (streaming) and plasmodium oscillations for ``mobiligence''\cite{mayne2015role}}}
\\\hline

\end{supertabular}
\end{flushleft}

\subsection{M{\"u}ller, di Primio and Lengeler's Table}

Bernd S. M{\"u}ller and Franco di Primio both of Institute for Autonomous intelligent Systems of German National Research Center for Information Technology and Joseph W. Lengeler of Deptartment of Biology/Chemistry, University of Osnabrück in Germany \cite{muller} have assembled the minimal cognitive abilities into the following list:
\begin{itemize}
\item Modifiable stimulus-response pathways [base of all behaviours]
\item Selective search
\item Perception
\item Memory
\item Expectations formation reaction [on the sudden deprivation of food]
\item Detecting the identity
\item Scrutiny [Counting in selective search]
\item Adaptation
\item Habituation
\item Learning
\item Cooperation [both for the individual living in a
population and the population as a whole]
\item Reconfigurability of the body [Synthesizing a situation-dependent or self-destruct of effectors and sensors; to avoid hunger]
\item Flexibility [tensegrity] This minimal cognition is amplified than M{\"u}ller et al contribuition. We aggregate the mechanosensory transduction by bioarchitecture \cite{ingber1997tensegrity} of filaments like vertebral column and wiring scaffold biophysics \cite{ingber2014tensegrity} for electrochemical connections. 
\end{itemize}

\begin{flushleft}
\tablefirsthead{}

\tabletail{%
\hline
\multicolumn{2}{|r|}{\small\sl continued on next page}\\
\hline}
\tablelasttail{}
\tablehead{%
\hline
\multicolumn{2}{|r|}{\small\sl continued from previous page}\\
\hline
}

\begin{supertabular}{m{2.1587598in}|m{3.7761598in}|}
\hline
\multicolumn{2}{|m{6.0136595in}|}{\centering {\textbf{Minimum Cognitive Principles}}}\\\hline
\multicolumn{1}{|m{2.1587598in}|}{{\textbf{M{\"u}ller, di Primio and Lengeler (2001)}}} &
{\textbf{{\it P. polycephalum} Behaviour}}\\\hline
\multicolumn{1}{|m{2.1587598in}|}{{\multirow{4}{2.1587598in}{Modifiable stimulus-response pathways [base of all behaviours]}}} &
{Minimum travel toward food \cite{nakagaki2000intelligence}}
\\\hhline{~-}
\multicolumn{1}{|m{2.1587598in}|}{} &
{Retroactivity negative \cite{nakagaki2000intelligence}}
\\\hhline{~-}
\multicolumn{1}{|m{2.1587598in}|}{} &
{ATP at the tip of plasmodium supervenes locomotive capacity \cite{takagi2007indecisive}}
\\\hhline{~-}
\multicolumn{1}{|m{2.1587598in}|}{} &
{Detection stimulus intensity differences through the Weber-Fechner Law \cite{mori2013cognition}}
\\\hline

\multicolumn{1}{|m{2.1587598in}|}{Selective search} &
{Rules for Biologically Inspired Design adaptive networks \cite{tero2010rules}}
\\\hhline{~-}
\multicolumn{1}{|m{2.1587598in}|}{\multirow{4}{*}{Selective search}} &
{Comparative evaluation standards for decision making food \cite{latty2011irrational}}
\\\hhline{~-}
\multicolumn{1}{|m{2.1587598in}|}{} &
{Precise speed decision making \cite{latty2011speed}}
\\\hhline{~-}
\multicolumn{1}{|m{2.1587598in}|}{} &
{Decision making, taking into account the variables of the key strengths of different foods \cite{reid2013amoeboid}}
\\\hhline{~-}
\multicolumn{1}{|m{2.1587598in}|}{} &
{Decision-making for solves the two-armed bandit \cite{reid2016decision}}
\\\hline

\multicolumn{1}{|m{2.1587598in}|}{{\multirow{2}{*}{Perception}}} &
{Barriers of maze \cite{nakagaki2000intelligence}}
\\\hhline{~-}
\multicolumn{1}{|m{2.1587598in}|}{} &
{Foraging sources \cite{nakagaki2000intelligence}}
\\\hline

\multicolumn{1}{|m{2.1587598in}|}{{\multirow{2}{*}{Memory}}} &
{Anticipation and memory periodic pulses \cite{saigusa2008amoebae}}
\\\hhline{~-}
\multicolumn{1}{|m{2.1587598in}|}{} &
{Using his trail for spatial memory \cite{reid2012slime}}

\\\hline
\multicolumn{1}{|m{2.1587598in}|}{{{Expectations formation reaction [on the sudden deprivation of food]}}} &
{Reaction-diffusion model for streaming oscillation patterns \cite{yamada2007dispersion}}
\\\hhline{~-}
\multicolumn{1}{|m{2.1587598in}|}{} &
{Distribution of nutrients \cite{dussutour2010amoeboid}}
\\\hline

\multicolumn{1}{|m{2.1587598in}|}{{\multirow{2}{*}{Detecting the identity}}} &
{Detection of changes in rates between separate paths to take parts \cite{reid2012slime}}
\\\hhline{~-}
\multicolumn{1}{|m{2.1587598in}|}{} &
{Minimum distance \cite{nakagaki2000intelligence}}
\\\hline

\multicolumn{1}{|m{2.1587598in}|}{\multirow{4}{*}{Scrutiny [in selective search]}} &
{Composition of a set of networks basic topologies \cite{nakagaki2004obtaining}}\\\hhline{~-}
\multicolumn{1}{|m{2.1587598in}|}{} &
{Rules for biologically inspired design adaptive networks \cite{tero2010rules}}\\\hhline{~-}
\multicolumn{1}{|m{2.1587598in}|}{} &
{Creating minimum transport networks \cite{latty2011structure}}\\\hhline{~-}
\multicolumn{1}{|m{2.1587598in}|}{} &
{Decision making and ``scatter'' or state of indecision between different attractors for foraging \cite{takagi2007indecisive}}
\\\hline

\multicolumn{1}{|m{2.1587598in}|}{{Adaptation}} &
{Morphodynamic changes that depend on the environment for motility \cite{takamatsu2009environment}}
\\\hhline{~-}
\multicolumn{1}{|m{2.1587598in}|}{\multirow{2}{*}{Adaptation}} &
{Rules for Biologically Inspired Design Adaptive Networks \cite{tero2010rules}}
\\\hhline{~-}
\multicolumn{1}{|m{2.1587598in}|}{} &
{The stimulus comes from attractants or repellants (external) provides the impetus for
morphological adaptation \cite{mayne2015role}. Maximizing internal flows by adapting patterns of contraction to size \cite{alim2017mechanism}.}
\\\hline

\multicolumn{1}{|m{2.1587598in}|}{{\multirow{2}{*}{Habituation [creating habits]}}} &
{Responsiveness decline and spontaneous recovery \cite{boisseau2016habituation}}
\\\hhline{~-}
\multicolumn{1}{|m{2.1587598in}|}{} &
{Habituation to be exposed to an innocuous repellent and transfer learning to another cell no exposed \cite{vogel2016direct}.}
\\\hline

\multicolumn{1}{|m{2.1587598in}|}{\multirow{2}{*}{Learning}} &
{The aversive response still occurred to another stimulus \cite{boisseau2016habituation}}
\\\hhline{~-}
\multicolumn{1}{|m{2.1587598in}|}{} &
{Associative learning to acquire a reversed thermotactic property, a new preference for the lower temperature \cite{shirakawa2011associative}.}
\\\hline

\multicolumn{1}{|m{2.1587598in}|}{{\multirow{4}{*}{Cooperation}}} &
{Production of fruiting bodies \cite{seifriz1931structure}\cite{seifriz1936protoplasm}\cite{seifriz1938recent}}
\\\hhline{~-}
\multicolumn{1}{|m{2.1587598in}|}{} &
{Interdependent collective irrationality of cooperative behaviour and violation of the principle of collective decision regularly supply \cite{latty2011speed}}
\\\hhline{~-}
\multicolumn{1}{|m{2.1587598in}|}{} &
{To leave chemicals in the environment, which are attractive to congeners to find food. Communication-based to cooperation and sociality \cite{vogel2015phenotypic}}
\\\hhline{~-}
\multicolumn{1}{|m{2.1587598in}|}{} &
{Fusion microplasmodia to form macroplasmodia in a percolation transition \cite{fessel2012physarum}}
\\\hline

\multicolumn{1}{|m{2.1587598in}|}{\multirow{2}{*}{Flexibility [tensegrity]}} &
{Unit microtubules for migrating towards the conductive cores plasmodium area \cite{ueda2000microtubule}}
\\\hhline{~-}
\multicolumn{1}{|m{2.1587598in}|}{} &
{Ratio of electric current flow and polarization associated with the actin-myosin and microtubules that coordinate the flow (streaming) of the plasmodium and its oscillations to the ``mobiligence'' \cite{mayne2015toward}}
\\\hline

\multicolumn{1}{|m{2.1587598in}|}{{{\multirow{2}{2.1587598in}{Body reconfigurability [Synthesising a situation-dependent or self-destruction of effectors and sensors; to avoid hunger]}}}} &
{Plasmodium to sclerotia \cite{seifriz1939materialistic}\cite{gehenio}\cite{jump1954studies}

}
\\\hhline{~-}
\multicolumn{1}{|m{2.1587598in}|}{} &
{Reaction-diffusion parameters under sol-gel conditions the plasmalemma \cite{takagi2007indecisive}

}
\\\hline

\multicolumn{1}{m{2.1754599in}}{~} &
\multicolumn{1}{m{1.8163599in}}{~}\\
\end{supertabular}
\end{flushleft}

\subsection{Gregory Bateson's Table}

Gregory Bateson's contribution to cognitive biology is that he sat that matter and energy are impregnated of circular processes of information of differences, creating "patterns that connect". His project was to explain the mind in terms of complexity and cybernetic organisation in the way that he was conceived it.

The six criteria that integrate the cognitive processes, or as Bateson calls it, mental processes are:

1. A mind is a set of interacting parts or components. It is an autopoietic red concept that is a network of interacting components.

2. The interaction of the parts of the mind is triggered by difference, and difference is a non-substantial phenomenon that is found neither in space nor time, the difference is related to neguentropy and entropy rather than to Energy.

3. The mental process requires collateral energy. In this criterion Bateson highlights the distinction between the ways in which living organisms and non-living systems interact with their environments. What Bateson exposes is that to describe the energy of life and to describe that of the forces and clashes of the non-living there should be a differentiated syntax.

4. The mental process requires circular (or more complex) processes of determination. Like Uexk{\"u}ll's Funktionkreises, biofeedback's Bertalanffy and Autopoiesis' Maturana and Varela.

5. In the mental process, the effects of difference must be seen as transformations (i.e. coded versions) of events that have preceded them.

6. The description and classification of these processes of transformation reveal a hierarchy of logical prototypes immanent in phenomena.

These criteria by Gregory Bateson on cognitive processes have been linked to the cognitive processes of Humberto Maturana.

\begin{flushleft}
\tablefirsthead{}
\tablelasttail{}
\tabletail{%
\hline
\multicolumn{2}{|r|}{\small\sl continued on next page}\\
\hline}
\tablelasttail{}
\tablehead{%
\hline
\multicolumn{2}{|r|}{\small\sl continued from previous page}\\
\hline
}
\begin{supertabular}{m{2.11566in}|m{3.91776in}|}
\hline
\multicolumn{2}{|m{6.1121597in}|}{\centering {\textbf{Mental/Cognitive processes}}}\\\hline
\multicolumn{1}{|m{2.11566in}|}{{\textbf{Gregory Bateson (1979)}}} &
{\textbf{{\it P. polycephalum} Behaviour}}\\\hline
\multicolumn{1}{|m{2.11566in}|}{{\multirow{5}{2.11566in}{A mind is an aggregate of interacting parts or components}}} &
{Interdependent collective irrationality of cooperative behaviour for decision making \cite{latty2011irrational}}
\\\hhline{~-}
\multicolumn{1}{|m{2.11566in}|}{} &
{Precise speeds and decision-making \cite{latty2011speed}}\\\hhline{~-}
\multicolumn{1}{|m{2.11566in}|}{} &
{Distribution of nutrients \cite{dussutour2010amoeboid}}\\\hhline{~-}
\multicolumn{1}{|m{2.11566in}|}{} &
{Using his trail for spatial memory \cite{reid2012slime}}\\\hhline{~-}
\multicolumn{1}{|m{2.11566in}|}{} &
{Association with actin filaments and microtubules myosin coordination of current flow (streaming) of the plasmodium and its oscillations to the ``mobiligence'' \cite{mayne2015toward}}\\\hline

\multicolumn{1}{|m{2.11566in}|}{\multirow{3}{2.11566in}{The interaction of the parts of mind is triggered by
difference}} &
{Using his trail for spatial memory \cite{reid2012slime}}\\\hhline{~-}
\multicolumn{1}{|m{2.11566in}|}{} &
{Collective cooperativeness in making food choices \cite{latty2011irrational}}\\\hhline{~-}
\multicolumn{1}{|m{2.11566in}|}{} &
{Detection stimulus intensity differences through the Weber-Fechner Law \cite{mori2013cognition}}\\\hline
\multicolumn{1}{|m{2.11566in}|}{{The mental process requires collateral energy}} &
{Distribution of nutrients \cite{dussutour2010amoeboid}}\\\hline
\multicolumn{1}{|m{2.11566in}|}{{Mental process requires circular (or more complex) determination processes}} &
{Reaction-diffusion model for streaming oscillation patterns \cite{yamada2007dispersion}}
\\\hline

\multicolumn{1}{|m{2.11566in}|}{{\multirow{6}{2.11566in}{In the mental process, the effects of the difference should be seen as transformations }}} &
{Internal communication: Reaction-diffusion model \cite{yamada2007dispersion}}
\\\hhline{~-}
\multicolumn{1}{|m{2.11566in}|}{} &
{Decision making and ``scattor'' or state of indecision between different attractors for foraging \cite{takagi2007indecisive}}
\\\hhline{~-}
\multicolumn{1}{|m{2.11566in}|}{} &
{Rules for biologically inspired design adaptive networks \cite{tero2010rules}}
\\\hhline{~-}
\multicolumn{1}{|m{2.11566in}|}{(i.e., coded versions) of events that have preceded} &
{Comparative assessment standards for making food choices \cite{latty2011irrational}}
\\\hhline{~-}
\multicolumn{1}{|m{2.11566in}|}{} &
{Precise speed and decision-making \cite{latty2011structure}}
\\\hhline{~-}
\multicolumn{1}{|m{2.11566in}|}{} &
{Decision-making for solves the two-armed bandit \cite{reid2016decision}}
\\\hline

\multicolumn{1}{|m{2.11566in}|}{The description and classification of these transformation} &
{Rules for biologically inspired design adaptive networks \cite{tero2010rules}}

\\\hline
\multicolumn{1}{|m{2.11566in}|}{{\multirow{2}{2.11566in} {Processes reveal a logical hierarchy of immanent phenomena prototypes}}} &
{Comparative assessment standards for making food choices and violation of the ``principle of Condorcet'' \cite{latty2011irrational}}\\\hhline{~-}
\multicolumn{1}{|m{2.11566in}|}{} &
{Precise speed decision making \cite{latty2011structure}}
\\\hline
\multicolumn{1}{m{2.1754599in}}{~} &

\multicolumn{1}{m{1.8163599in}}{~}\\
\end{supertabular}
\end{flushleft}

\subsection{Maturana's Table}

While Bateson worked from the scientific intuition forged from the observational method, typical of psychoanalysis, Maturana made it from the cybernetics itself with an innovative language. To paraphrase Paul Dell \cite{dell1985understanding}, Maturana contains the ontology that Bateson did not develop. While Bateson exposes his ideas in a sort of ``cosmology'' of biology for a cybernetic epistemology, Maturana realizes an ontology of biology in an epistemology that conjugates with the ideas of Bateson. Maturana sums up his analogy with Bateson's ideas in these three points previously implicated with Bateson's:
\begin{enumerate}
\item An autopoietic network is a network of interacting components. Here he conjugates with Bateson insofar as the mind is an aggregate of parts or interacting components. In other words, mind/cognition and autopoietic network support two plausible homeomorphic stages (not identical in themselves, as observed from the incorrect tautology between life and cognition). While in Bateson the interaction of parts of the mind is triggered by difference, for Maturana the difference is a non-substantial phenomenon that is found neither in space nor in time. The difference is rather related to neguentropy and entropy, rather than to energy \cite{brier2001cybersemiotics} \cite{brier2008bateson}.
\item The response of an organism requires structural coupling and non-linear patterns. Something that fits the idea of Bateson, in which a mental process requires a collateral energy.
\item The characterization of a living system, in terms of nonlinear patterns, infers autopoiesis. Nonlinear causality generates dissipative structures. In this case it follows the point at which Bateson explains that a mental/cognitive process requires circular (or more complex) processes of determination. Autopoietic systems require nonlinear patterns such as the reaction-diffusion equations, and serve as a non-linear pattern for uncoupled oscillations in the  Physarumm. Likewise, the concentration factors of multiple cells of {\it Dictyostelium discoideum}, or of {\it Dictyostelium mucoroides} --- already observed by Uexk{\"u}ll \cite{von1940bedeutungslehre} --- by the cAMP generate patterns of dissipative structures, excitation patterns, of Belousov-Zhabotinsky medium. These, among other factors, such as the role of Ca$^{2+}$ in actin-myosin in the streaming of Physarum protoplasm are part of what are considered as adaptive regulatory factors, fundamental for the generation of minimal cognitive processes. Let us see how this theory is deduced, from the role of biological autonomy.
\end{enumerate}

\begin{flushleft}
\tabletail{%
\hline
\multicolumn{2}{|r|}{\small\sl continued on next page}\\
\hline}
\tablelasttail{}
\tablehead{%
\hline
\multicolumn{2}{|r|}{\small\sl continued from previous page}\\
\hline
}
\tablelasttail{}
\begin{supertabular}{m{2.07476in}|m{3.9594598in}|}
\hline
\multicolumn{2}{|m{6.11296in}|}{\centering {\textbf{Mental/Cognitive processes}}}\\\hline
\multicolumn{1}{|m{2.07476in}|}{{\textbf{Humberto Maturana (1970)}}} &
{\textbf{{\it P. polycephalum} Behaviour}}\\\hline
\multicolumn{1}{|m{2.07476in}|}{{\multirow{2}{2.11566in}{An autopoietic network is a network of interacting components}}} &
{Interdependent behaviour irrationality of collective decision making \cite{latty2011irrational}}\\\hhline{~-}
\multicolumn{1}{|m{2.07476in}|}{} &
{Precise speeds and decision-making \cite{latty2011structure}}\\\hhline{~-}
\multicolumn{1}{|m{2.07476in}|}{} &
{Distribution of nutrients \cite{dussutour2010amoeboid}}\\\hhline{~-}
\multicolumn{1}{|m{2.07476in}|}{} &
{Using his trail for spatial memory \cite{reid2012slime}}\\\hhline{~-}
\multicolumn{1}{|m{2.07476in}|}{} &
{Relationship of flow of electric current and polarization associated with filaments and myosin microtubules, in turn coordinate current flow (streaming) and plasmodium oscillations for ``mobiligence'' \cite{mayne2015role}}
\\\hline

\multicolumn{1}{|m{2.07476in}|}{{\multirow{2}{2.07476in}{The response of an organism requires structural coupling and
nonlinear patterns}}} &
{Anticipatory mechanisms underlying behaviour from the perspective of nonlinear dynamical systems \cite{nakagaki2004obtaining}}\\\hhline{~-}
\multicolumn{1}{|m{2.07476in}|}{} &
{Decision making and ``scatter'' or state of indecision between
different attractors for foraging \cite{takagi2007indecisive}}\\\hhline{~-}

\multicolumn{1}{|m{2.07476in}|}{\multirow{4}{2.07476in}{}} &
{Arboreal behaviour patterns similar to excitation waves in Belousov-Zhabotinsky medium  \cite{adamatzky2009reaction}\cite{adamatzky2009if}}\\\hhline{~-}
\multicolumn{1}{|m{2.07476in}|}{} &
{Rules for biologically inspired design adaptive networks \cite{tero2010rules}}\\\hhline{~-}
\multicolumn{1}{|m{2.07476in}|}{} &
{Detection stimulus intensity differences through the Weber-Fechner Law \cite{mori2013cognition}}\\\hhline{~-}
\multicolumn{1}{|m{2.07476in}|}{} &
{The stimulus comes from attractants or repellants (emiulxternal) provides the impetus for morphological adaptation \cite{mayne2015role}}\\\hline
\multicolumn{1}{|m{2.07476in}|}{{\multirow{4}{2.11566in}{The characterization of a living system, in terms of nonlinear patterns Autopoiesis inferred. The non-linear causality generates dissipative structures.}}} &
{Anticipatory mechanisms underlying behaviour from the perspective of nonlinear dynamical systems \cite{nakagaki2004obtaining}}\\\hhline{~-}
\multicolumn{1}{|m{2.07476in}|}{} &
{Reaction-diffusion  model for streaming oscillation patterns \cite{yamada2007dispersion}}\\\hhline{~-}
\multicolumn{1}{|m{2.07476in}|}{} &
{Decision making and ``scatter'' or state of indecision between different attractors for foraging \cite{takagi2007indecisive}}\\\hhline{~-}
\multicolumn{1}{|m{2.07476in}|}{} &
{Arboreal behaviour patterns similar waves in Belousov-Zhabotinsky system \cite{adamatzky2009reaction}\cite{adamatzky2009if}}\\\hline
\multicolumn{1}{m{2.07476in}}{~
} &
\multicolumn{1}{m{3.9594598in}}{~
}\\
\end{supertabular}
\end{flushleft}

\section{Emerging Sources of Cellular Levels of Sentience and Consciousness}

Consciousness is emerging as a basic and inherent property of biological organisms which is relevant for their survival and evolution \cite{mashour2013evolution} \cite{barron2016insects} \cite{baluvska2016understanding} \cite{calvo2017plants}. Importantly in this respect, plants and several unicellular organisms generate endogenous anesthetics any time they are wounded or stressed \citet{baluvska2009deep} \cite{tsuchiya2017anesthetic}.
Hypothetical basic unit of consciousness in multicellular organisms, such as humans, non-human animals and plants \cite{griffin2004new} \cite{baluvska2009deep} \cite{trewavas2011ubiquity} \cite{gardiner2012insights} \cite{gardiner2015subcellular} \cite{barlow2015natural} \cite{barron2016insects} \cite{baluvska2016understanding} \cite{calvo2017plants}, might be represented by cellular and subcellular levels of consciousness \cite{margulis2000microbial} \cite{margulis2001conscious}. There are at least three possible sources of sentience and consciousness (understood as a gradual self-mapping tool) at the cellular and subcellular levels.

It is important to realize that ion channel and neurotransmitter hardware, which underlie the software of cognition in brains, are evolutionarily ancient and pre-date multicellularity \cite{levin2006minds,liebeskind2011evolution}.  For a review of cognitive approaches to the decision-making and memory in non-neural metazoan cells and tissues in the context of embryogenesis, regeneration, and cancer suppression see \cite{baluvska2016having,pezzulo2015re,mathews2017gap}.

The excitable membranes with critical proteins embedded especially in highly ordered lipid rafts, vibrating and excitable microtubules and actin filaments, as well as biological quasicrystals based on the five-fold symmetry.

Possible relevances of biological excitable membranes for cellular consciousness is supported by their sensitivites to diverse anaesthetics inducing loss of consciousness in humans, as well as loss of responsiveness in animals and plants \citet{perouansky2012quest} \cite{mashour2013evolution} \cite{gremiaux2014plant}. It is important to be aware that electrically-active membranes evolved early in the biological evolution \cite{wayne1994excitability} \citet{johnson2002action} \cite{eckenhoff2008can} \cite{la2012effects} \cite{cook2014membrane} \cite{brunet2016damage} and are present even in prokaryotic organisms and eukaryotic organelles of endosymbiotic origin \cite{masi2015electrical} \cite{prindle2015ion} \cite{catterall2015deciphering} \cite{baluvska2016understanding}. Our recent report reveal that the sensitivity of plant movements and behaviour to local and general anesthetics is linked to membranes, action potentials, and to endocytic vesicle recycling~\cite{yokawa2017}.
Another possible source of sentience  at the cellular and subcellular levels is the dynamic cytoskeleton. Especially microtubules are discussed as important in this respect \cite{hameroff2002conduction} \cite{gardiner2012insights} \cite{gardiner2015subcellular} \cite{barlow2015natural}, representing quantum channels related to consciousness \cite{ja2015anesthetics} \cite{tonello2015possible} and terahertz oscillations in tubulin were reported to be affected by exposures to anesthetics \cite{craddock2017anesthetic}. Besides microtubules, also the actin filaments behave as excitable medium which transports ionic waves and mediates eukaryotic chemotaxis in response to diverse gradients \cite{tuszynski2004ionic} \cite{iglesias2012biased} \cite{tang2014evolutionarily} \cite{van2017coupled}. Actin cytoskeleton supports lipid rafts, which are ordered domains of biological membranes particularly sensitive to anesthetics \cite{morrow2005flotillins} \cite{tsuchiya2010local} \cite{bandeiras2013anesthetics} \cite{weinrich2013xenon}.

There are also indications that special proteins or assembly of proteins, especially those having five-fold symmetry and quasicrystals properties, seem to be relevant for the cellular and subcellular levels of sentience and consciousness \cite{gardiner2012insights} \cite{gardiner2015subcellular}.

\section{Proto-consciousness and Morgan's Canon}

Up to date, consciousness has been studied in humans and mainly focused into finding neural correlates of such pervasive phenomena. Because of the complexity of its study, it has been called ``The hard problem'' \cite{vallverdu}. Despite of the evident and implicit epistemical challenges, this problem is also related to the necessity of a theory about the functional meaning of consciousness, from a cognitive-information perspective. Consciousness, or any related self-awareness checking, is essential for living systems to enjoy their agency via Sense of Self. This allows them to act in their own interest: with memories, predictions, learning, decisions etc. Thus, we must remark that Consciousness, Self-awareness and Sense of Self allow living systems to recognize and predict patters of external (and internal!) environmental challenges, something which implies better adaptation, survival and evolution rates. This self-awareness can be understood as a genome adaptive mechanism \cite{bacterial} necessary for molecular cooperation and, at an evolutionary perspective, basic for adaptive mutagenesis.

A sense of Self is based on consciousness and anesthetics switch-off both. Living systems under anesthesia will not survive for very long as they lack Sense of Self and cognitive agency.

At least in humans, consciousness can be defined as the experiential awareness of the own existence, but it has demonstrated not to be the hard control system of the whole sum of cognitive processes, but instead of it, consciousness must be defined as a super-level mechanism of punctual data integration for specific decision-taking actions. Therefore we can trace early mechanisms of data integration for evaluation purposes in slime mould that would suggest a proto-consciousness emergence.  Proto-Consciousness is essential for living systems – again allows organisms to understand complex context of their environmental situation. They can acts agents successful in (a) finding food; (b) avoiding predators; (c) finding mating partners, etc. It is related to the notion expressed by neurologists as ``proto-qualia'' \cite{llinas2001vortex}, as the mechanism to differentiate the sensivity of an specific perception. This is mechanistically explained by Weber-Fechner law \cite{castro2011principles}. Using \cite{llinas1990intrinsic,llinas2014intrinsic}, we can suggest as plausible to consider both the Physarum and CNS mammarial cells (as neurons) has as identic isotype of tubulin for build microtubules and centrioles (for cellular division or cilliar composition). But while these in Physarum are alpha-tubulins, in mammarial cells are beta-tubulins. These changes in ``peanut'' tubulin structure create in both different electrophysiological properties. It is very interesting to follow these changes in order to find a link between brainless cells and brain cells behaviours and biomechanical principles.

Following such a naturalistic approach to consciousness explained by \cite{vallverdu2017}, we will explore the validity and usefulness in this debate of \cite{tononi2016}, and \cite{thagard2014}. This is at the same time a process related to a naturalistic statistical approach to data processing, also called 'natural statistics' \cite{vallverdu2015bayesians} and that could provide a Bayesian approach to consciousness \cite{gunji2017} which fits perfectly with Kolmogorov algorithmic information theory \cite{kolmogorov1968logical}, which provides a solid way to deal with the quantity of information in a system, and at the end, and in some fundamental way, living systems can be considered information processing entities for reproduction purposes. It is also interesting to note that Bayesian models fit well with slime mould information processing \cite{schon2014physarum, miyaji2008physarum}.

At the same time, we strongly defend that the notion of proto-consciousness can provide us a naturalistic, evolutionary and bottom up mechanism of explaining consciousness. As Conwy Lloyd Morgan wrote in his ``An Introduction to Comparative Psychology'' in 1903~\cite{morgan1903introduction}: 
\begin{quote}
``In no case is an animal activity to be interpreted in terms of higher psychological processes if it can be fairly interpreted in terms of processes which stand lower in the scale of psychological evolution and development.''
\end{quote}

For evident reasons, at this low biological level we cannot look for neural correlates, instead of biochemical correlates of (proto)consciousness. 

Our starting point is simple: slime mould computes information. And there are some basic mechanisms to compute it that allow us to talk about proto-consciousness in  brainless living systems,  easily identifiable in slime mould. We will see it into the next section.

\section{The Computing Slime Mould as Kolmogorov-Uspensky Biomachine}

Late 1940s Andrei Kolmogorov published his seminal paper ``On the concept of algorithm'' \cite{kolmogorov1953concept} where he introduced a concept of an abstract machine -- later called Kolmogovor-Uspensky machine -- defined on a dynamically changing graph structure, a computational process on a finite indirected graph with distinctly labelled nodes. A computational process travels on the graph, activates nodes and removes and adds edges~\citet{uspensky1992kolmogorov}.  Basic operations of the machine include: selecting of a node, specifying of node's neighbourhood, adding and/or removing nodes and edges. A program for the machine specifies how to replace the neighbourhood of an active node with a new neighbourhood,
depending on the labels of edges connected to the active node and the labels of the nodes in proximity to the active node~\cite{blass2003abstract}. Gurevich~\cite{gurevich_1988} suggested that the edge of the `Kolmogorov complex' is not only  informational but also a physical entity and reflects the physical proximity of the nodes (e.g. even in three-dimensional space the number of neighbors of each  node is polynomially bounded). But most revolutionary feature of the Kolmogovor-Uspensky machine is its `physical nature'; as Andreas Blass and Yuri Gurevich~\cite{blass2003abstract} stated: 
\begin{quote}
``Turing machines formalize computation as it is performed by a human. Kolmogorov machines formalize computation as it performed by a physical process.''
\end{quote}
In \cite{adamatzky2007physarum} we demonstrated in experimental laboratory conditions that Physarum can be seen as living implementation of the Kolmogorov-Uspensky machine, where nodes are represented by sources of nutrients and branching points of the network of protoplasmic tubes, programmed by spatial gradients of attractants and repellents. Each part of Physarum `reads' and reactors to quanta of information written in its environment by acting depending on on the state of its neighbourhood. 

Like a Kolmogorov-Uspensky machine embodies data and program in topology of its graph the slime mould can compute several sets of data without the necessity of a high level integration, or as Andrei Kolmogorov wrote \cite{kolmogorov1958definition}: 
\begin{quote}
``... the schema for computing the value of a partially recursive function may not
be directly given in the form of algorithm. If one develops this computation in the
form of an algorithmic process ... then, through this, one automatically obtains a
certain algorithm ...''
\end{quote}
The point here is to justify bio-mechanically how different data and different kinds of actions are evaluated and coordinated within a slime mould. This could provide us the key  for understanding the existence of a proto-consciousness.  Although we have listed a long list of cognitive patterns, these processes are not just mechanistic/automated answers to the environment, but imply a certain evaluation and decision process. This upper level is not a generally requested task involved into all cognitive processes of slime mould, but just a small part of them. From an informational perspective, the process of data binding or integration requires from a mechanism that when is found in other cognitive systems is often referred as `consciousness'. This can be inferred from collective behaviour of slime mould, clearly intelligent, and that allows us to explain the hypothetical model that captures the mechanistic way of transferring informational states from one functional state to another one. That is, \textit{{the path from automation to decision}}. In the next section we will explore the functional properties of this brainless consciousness mechanism of slime mould.

\section{Slime mould complexity and brainless information integration system}

It is beyond any doubt that slime mould do not have brain nor any nervous central system. Despite of this fact, it is also true that regardless of this fact, they are able to perform incredible cognitive tasks  that go beyond automated chemical responses to the environment. For that reason, we defend the existence of a proto-consciousness mechanism in slime mould colonies. Without it we could not explain very complex responses and adaptations to the organism. 

Following the formalization of the Integrated Information Theory of consciousness (henceforth, IIT)  \cite{oizumi2014phenomenology}, information is a 
\begin{quote}
``$\ldots$ set of elements can be conscious only if its mechanisms specify a set of ``differences that make a difference'' to the set --– i.e. a conceptual structure. A conceptual structure is a constellation of points in concept space, where each axis is a possible past/future state of the set of elements, and each point is a concept specifying differences that make a difference within the set. The higher the number of different concepts and their max value, the higher the conceptual information CI that specifies a particular constellation and distinguishes it from other possible constellations.''
\end{quote}

We have provided in previous sections several examples about how slime mould execute integration and exclusion informational decisions. Composition is also an accomplished axiom of IIT because slime mould decide according to the environmental changes that modify substantially their behaviour.

Slime mould is an ideal minimal organism to test for consciouness in particular using ideas from IIT given its many computational and informational properties and because IIT allows a continuous spectra of  degrees rather than a binary answer to the question of consciousness.

In \cite{maguire} and \cite{gauvritzenil} it is shown how IIT is deeply and, in a formal setting, fundamentally connected to the concept of algorithmic complexity and data compression and in \cite{zenilmarshall} how algorithmic probability---inversely related to algorithmic complexity by a formal theorem---may explain  aspects of biological evolution. More recently, it has been shown that such foundations have the ability to explain a wide range of evolutionary phenomenology that would remain unexplained when making traditional assumptions related to the role of random mutation in natural selection and how organisms rather harness the highly structured nature of ecosystems \cite{algoevo}.

In \cite{zenilmarshalltegner} it is also shown how experimental behavioural observations can be explained  by algorithmic complexity (as opposed to statistical tools) and help explain behavioural phenomenology in areas such as working memory \cite{Chekaf2015ChunkingIW,chunking2}, psycometrics \cite{gauvritprimer,gauvritfriendly}, visual probabilistic reasoning \cite{gauvritcognition}  and cognitive structures of cultural nature\cite{gauvritcognition} able to quantify a multiple range of cognitive processes that were more difficult, if not impossible (both for practical and fundamental reasons), to characterize using conventional (non-algorithmic) statistical tools. In \cite{zenilentropy} it is shown how organisms can harness the algorithmic structure of the natural environment mirroring the non-random structure of the physical world and using the same tools it was shown how algorithmic complexity can be used to validate behavioural results of animals \cite{zenilmarshalltegner}, including foraging communication by ants, flight patterns of fruit flies, and tactical deception and competition strategies in rodents. 

In fact, the only application of IIT to this date \cite{casali} completely relied on the concept of algorithmic complexity as a foundation of their proposed numerical measure using lossless compression applied on spatiotemporal patterns of electrocortical responses of humans using Lempel-Ziv where it was found to reliably discriminate levels of consciousness during wakefulness, sleep, anaesthesia and minimally conscious states even though such approach is closer to a measure of entropy and entropy has been shown to be extremely limited as a computable measure \cite{zenilphysrev,maguire}.  The nontrivial behaviour of slime mould and its intricated relationship of awareness and interaction with the environment suggests that one can device behavioural time series of slime mould behaviour and apply all these powerful tools and methods of algorithmic cognition \cite{zenil} to determine the complexity of its behavioural capabilities and shed light on measures of integrated information suggesting some degree of consciousness in an experimental/numerical setup as an obvious experiment.

%\begin{figure}
%\centering
%\subfigure[]{\includegraphics[width=0.49\textwidth]{PhysarumBrain}}
%\subfigure[]{\includegraphics[width=0.49\textwidth]{PhysarumBrainElectrodes}}
%\subfigure[]{\includegraphics[width=0.99\textwidth]{PhysasrumBrainREcroding}}
%\caption{Physarum brain. (a) slime mould growing on an agar brain. (b) positions of electrodes. (c) electrical potential recorded during 50 hours.}
%\label{fig:physarumbrain}
%\end{figure}

%\begin{figure}
%\centering
%\subfigure[]{\includegraphics[width=0.7\textwidth]{TEF_2}}
%\subfigure[]{\includegraphics[width=0.7\textwidth]{TEF_3}}
%\caption{Response of slime mould to trifluoroethane (Sigma Aldrich, UK). Electrical potential difference between two sites of 10~mm long protoplasmic tube was measured using aluminium electrodes, amplified and digitised with ADC-20 (Pico Technology, UK). (a)~5~$\mu$L of trifluoroethane applied to a 5~mm$\times$5~mm piece of filter paper placed in the Petri dish with slime mould. (b)~25~$\mu$L applied.}
%\label{fig:TEF}
%\end{figure}
%let us speculate on what would happen if we had a slime mould instead of nervous system in our brain (Fig.~\ref{fig:physarumbrain}) anaesthesis (Fig.~\ref{fig:TEF}). 

\section{Concluding remarks}

So we conclude chemotaxis of myxomycetes  represents a clear comparative example of how certain fundamental capabilities for the origin of cognition may arise in the minimum living systems, and how this is possible only through the action of regulatory mechanisms. It is only in the presence of regulation that specific disturbances acquire a meaning for the system. This becomes, thus, a biosemiotic foundation of the basics of cognitive minimum principles by which emerging regulatory system factors acquire those minimum principles offered in cognitive biology \cite{maturana, maturela, bateson, lyon,lyon2, muller}. 

Also we have seen that experiments with anesthesia in plants are an outstanding framework to establish conditions of possibility to bridge sentience with a proto-consciousness \cite{yokawa2017}. Is necessary to review this experimental task with slime molds -as Seifriz made on for streaming behaviour studies \cite{seifriz1941theory}- and so to realize that ion channel and neurotransmitter hardware, which underlie the software of cognition in brains, are evolutionarily ancient and pre-date multicellularity.

According to the provided data and analysis, slime mould provides an insightful example  of a biosemiotic entity able to perform cognitive tasks and to explain the first steps from mechanistic automation to decision, as well as of coordination and cooperation, and also an ideal testbed for consciousness as a minimal conscious biological organism.

Physarum exemplifies embedding cognition and computation:  the slime mould perceives its world in parallel, process the information perceived concurrently, makes decisions in a decentralised manner and represents the decision, or results of the computation, in spatially distributed configuration of its protoplasmic tubes; the tubes configuration per se might act a program, similarly to Kolmogorov-Uspensky machine, which determines how the information perceived will be processed.

Finally, we have introduced Morgan's canon as a necessary epistemological condition to eradicate the anthropomorphization of the sense of cognition, as well as this bridge to consciousness.

\section{Acknowledgements}

The part of this work is performed according to the Russian Government Program of Competitive Growth of Kazan Federal University and by the subsidy allocated to Kazan Federal University for the state assignment in the sphere of s scientific activities number 1.4539.2017/8.9.

\newpage

\bigskip

\bigskip

\bigskip

%% The Appendices part is started with the command \appendix;
%% appendix sections are then done as normal sections
%% \appendix

%% \section{}
%% \label{}

%% References
%%
%% Following citation commands can be used in the body text:
%% Usage of \cite is as follows:
%%   \cite{key}          ==>>  [#]
%%   \cite[chap. 2]{key} ==>>  [#, chap. 2]
%%   \citet{key}         ==>>  Author [#]

%% References with bibTeX database:

\bibliographystyle{model1-num-names}
%\bibliographystyle{plain}
%\bibliography{biblbio} 

% \begin{thebibliography}{00}

%% \bibitem must have the following form:
%%   \bibitem{key}...
%%

% \bibitem{}

% \end{thebibliography}

\end{document}